\pdfoutput=1
\documentclass[a4paper,11pt]{article}
\usepackage[utf8]{inputenc}
\usepackage{amsmath}
\usepackage{graphicx}
\usepackage{url}
\usepackage[mathscr]{eucal}

\usepackage[labelfont=bf]{caption}
\usepackage{comment}
\usepackage{outlines}
\usepackage{subcaption}
\usepackage{enumitem}
\usepackage{lineno}
\usepackage{jinstpub}

\title{Characterizing the energy resolution of the MicroBooNE LArTPC at the MeV scale using monoenergetic features of $^{\mathbf{208}}$Tl decays}
\collaboration{MicroBooNE Collaboration}

\abstract{A detailed understanding of the capabilities and fidelity of low-energy reconstruction is crucial for taking advantage of MeV-scale neutrino physics opportunities in liquid argon time projection chambers (LArTPCs). This study presents a measurement of the resolution of reconstructed energy in the MicroBooNE LArTPC at $\approx 1.5$ MeV. The characterization is performed using monoenergetic signals generated by 
$2.614$ MeV $\gamma$-rays from $^{208}$Tl decays undergoing pair production in the detector. The resolution is found to be ($7.52 \pm 0.78 \text{ (stat)} \pm 0.92 \text{ (syst)}$)\%. This value is consistent with the MicroBooNE simulation prediction of ($9.70 \pm 0.65$\text{ (stat)})\% at the $1.6 \sigma$ level. This study represents the first ever measurement of LArTPC energy resolution at the MeV scale and provides a pathway for monoenergetic energy calibrations in future experiments using LArTPC detectors.
  
}


\begin{document}

\author[mm]{P.~Abratenko}
\author[n]{D.~Andrade~Aldana}
\author[ll]{J.~Asaadi}
\author[kk]{A.~Ashkenazi}
\author[l]{S.~Balasubramanian}
\author[l]{B.~Baller}

\author[dd]{A.~Barnard}
\author[dd]{G.~Barr}
\author[dd]{D.~Barrow}

\author[z]{J.~Barrow}
% ...

% ...
\affiliation[l]{Fermi National Accelerator Laboratory (FNAL), Batavia, IL 60510, USA}

% ...
\affiliation[n]{Illinois Institute of Technology (IIT), Chicago, IL 60616, USA}
% ...
\affiliation[z]{University of Minnesota, Minneapolis, MN, 55455, USA}
% ...
\affiliation[dd]{University of Oxford, Oxford OX1 3RH, United Kingdom}
% ...
\affiliation[kk]{Tel Aviv University, Tel Aviv, Israel, 69978}
\affiliation[ll]{University of Texas, Arlington, TX, 76019, USA}
\affiliation[mm]{Tufts University, Medford, MA, 02155, USA}
% ...

  \emailAdd{microboone\_info@fnal.gov}

\maketitle

\section{Introduction} \label{sec:intro}

The MeV energy scale holds exciting opportunities for neutrino physics using liquid argon time projection chambers (LArTPCs), the only large neutrino detector technology capable of delivering low energy thresholds alongside millimeter spatial resolution. For instance, the physics scope of GeV-scale accelerator neutrino experiments may be expanded to include solar and supernova neutrino detection \cite{Solar, supernova1, supernova2}, as well as new or enhanced searches for physics beyond the Standard Model (BSM)~\cite{ArgoNeuT:2019ckq,Harnik:2019zee, sterile, hps, hnl}. Sensitivity to MeV energies can also enable reconstruction of previously-invisible low-activity $\nu$-Ar vertices, final-state neutron identification \cite{neutron, neutron1, argoneut, neutron4}, and new detector response calibration \cite{calibration}.  Refs. \cite{MeV_opportunities, list_1, list_2} provide an overview of low-energy physics possibilities in neutrino experiments using LArTPCs (or `neutrino LArTPCs'). 

LArTPCs can sense and characterize energy deposits by charged particles in the form of ionization charge and scintillation light. While both are exploited in LArTPC-based dark matter experiments~\cite{Benetti:2007cd,DarkSide:2018bpj,DarkSide-20k:2017zyg}, neutrino LArTPCs have so far relied on ionization charge signals only (although including light has been shown to benefit low-energy calorimetry \cite{lariat_light}). In single-phase neutrino LArTPCs, an electric field drifts ionization electrons toward a charge collection system, most often composed of wire planes. The spacing of the sensing wires sets the mm-scale spatial resolution, and the low-noise readout electronics allow for sub-MeV energy detection thresholds \cite{argoneut, Radon}. 

MeV-scale energy depositions in LArTPCs spread over just a few wires on each readout plane, as the ionization clouds from low-energy charged particles in liquid argon have spatial extents on the order of the wire spacing. These small-featured topologies have required the development of a new object in signal reconstruction called a `blip', which is described in detail in \mbox{ref. \cite{Radon}}. Blips are high-level three-dimensional objects reconstructed from charge depositions on multiple planes, similar to tracks and electromagnetic showers. While these two latter topologies, produced by particles with $\approx0.01$--$10$ GeV energies, can extend for many centimeters or meters, blips are compact and isolated.

Efforts to leverage the low-energy capabilities of LArTPCs are relatively recent. Pioneering studies identified $\approx5$--$50$ MeV Michel electrons from cosmic muon decay in ICARUS and MicroBooNE \cite{Icarus, uboone_pioneer}. These were followed by ArgoNeuT's detection of blips from $\nu$-Ar final-state photons and neutrons \cite{argoneut}. LArIAT measured blips from final-state products of pion and muon nuclear capture at rest, providing a new method of pion-muon discrimination \cite{miguel}. MicroBooNE, after preliminary low-energy demonstrations \cite{mev_thesis, uboone_note, uboone_note2}, produced several analyses measuring MeV-scale radiogenic activity in the detector \cite{1stRadon, Radon, Diego}.

In accomplishing these low-energy physics measurements, LArTPC-based experiments have demonstrated key aspects of MeV-scale detector response. MicroBooNE's radioisotope studies validated a lowered-threshold reconstruction algorithm that achieves high blip identification efficiency well below 1~MeV \cite{Radon}.  
The most recent of these also determined a ($3.1 \pm 0.2 \text{ (stat) } \pm 1.2 \text{ (syst)}$)\% energy scale shift between data and simulation at the MeV scale \cite{Diego}. Additionally, \mbox{ref. \cite{Diego}} reported a new particle identification metric for blips based on size-to-energy comparisons that fully exploits the detector's spatial resolution. 

Even as the above-mentioned  signal identification thresholds and energy scale shift have been demonstrated with experimental data, the only LArTPC MeV-scale energy resolution characterizations so far have been obtained from simulation \cite{Radon,supernova1}. The energy resolution of calorimetric detectors can be parametrized by

\begin{equation}
    \frac{\delta E}{E}=\frac{a_0}{E\text{[MeV]}} \oplus \frac{a_1}{\sqrt{E\text{[MeV]}}} \oplus b \text{,}
    \label{eq:res}
\end{equation}

\noindent where $a_0$, $a_1$, and $b$ account for the electronic noise, counting statistics, and reconstruction-related systematic effects, respectively~\cite{formula}.  In \mbox{ref. \cite{Radon}}, equation~(\ref{eq:res}) was used to fit the results from a sample of electrons simulated uniformly throughout the MicroBooNE LArTPC, resulting in \mbox{$a_0 = (3.1 \pm 0.2)$\%}, $a_1=(6.4\pm0.2)$\%, and $b = (7.30 \pm 0.05)$\%.  This fit predicts the resolution of energy reconstruction in the MicroBooNE TPC to vary from roughly $10$\% at $1$ MeV to $8$\% at \mbox{$5$ MeV}.

In this study, we present the first MeV-scale energy resolution measurement in a neutrino LArTPC, obtained from the products of radioactive $^{208}$Tl decays inside MicroBooNE. The demonstrated technique and resolution outcome are highly relevant to current and future LArTPC neutrino experiments, such as Fermilab's SBN Program \cite{sbn, list_2}, ProtoDUNE \cite{protodune}, and DUNE \cite{dune}.
Section \ref{sec:tl208_LAr} offers an overview of $^{208}$Tl decay $\gamma$-ray production and interaction in liquid argon detectors, whereas section \ref{sec:tl208_uboone} examines the specific case of MicroBooNE. In section \ref{sec:blips}, we describe the datasets employed in this analysis and the blip reconstruction algorithms applied. Finally, section \ref{sec:selection} details the signal selection and background treatment, and section \ref{sec:analysis} reports the data fitting and statistical analysis of the results.

\section{$^{208}$Tl decay $\gamma$-rays in liquid argon} \label{sec:tl208_LAr}

Decay products of $^{208}$Tl are often observed in particle detectors due to the trace presence of its ancestor nuclides $^{232}$Th (in solids) or $^{220}$Rn (in gases or liquids).  
Radon activity in gas- and liquid-phase purified argon is expected to be relatively low --- while $^{220}$Rn concentrations have not been studied for MicroBooNE LAr, $^{222}$Rn specific activities are $<0.35$ mBq/kg \cite{Radon}. Still, $^{208}$Tl decay radiation can be observed from the presence of $^{232}$Th in detector structural materials, along with $^{238}$U and $^{40}$K. $^{208}$Tl $\beta$-decays ($Q_{\beta}=5.00$ MeV; $3.05$ minute half-life) to various excited states of $^{208}$Pb, yielding a spectrum of coincidentally emitted $\gamma$-rays. The most prominent of these is a $2.614$ MeV $\gamma$-ray produced in $>99$\% of $^{208}$Tl decays \cite{NuDat}, whose high energy and intensity make it measurable in many large particle detectors with little interference from other radiogenic features \cite{daya_bay, borexino, exo-200, prospect}. From here onward, we focus on this single $\gamma$-ray line of $^{208}$Tl decays.

$2.614$ MeV $\gamma$-ray interactions in liquid argon consist mostly of Compton scattering ($94.6$\%), but pair production contributes substantially as well ($5.24$\%) \cite{NIST}. The cross sections of these channels are displayed in table \ref{tab:cross-sections}, which also includes coherent scattering and photoelectric absorption, both of which are practically negligible. The total $\gamma$-ray cross section at this energy is $2.38$ b/atom, equivalent to an attenuation length of $19.9$ cm, with an assumed LAr density of $1.4$ g/cm$^3$. 

\begin{table}[h]
    \centering
    \caption{Cross sections ($\sigma$) of all possible $2.614$ MeV $\gamma$-ray interactions in LAr and probability of undergoing each interaction mode \cite{NIST}. \vspace{0.5\baselineskip}}
    \begin{tabular}{c|c|c}
        \textbf{Process} & \textbf{$\sigma$} (b/atom) & \textbf{Fraction} (\%) \\
        \hline
        \hline
         Coherent scattering & $2.24 \times 10^{-3}$& $0.09$ \\
         Compton scattering & $2.26$ & $94.6$ \\
         Photoelectric absorption & $ 1.09 \times 10^{-3}$& $0.05$ \\
         Pair production & $0.125$ & $5.24$ \\
         \hline
    \end{tabular}
    \label{tab:cross-sections}
\end{table}

\begin{figure}[h]
    \centering
    \includegraphics[width=0.8\linewidth]{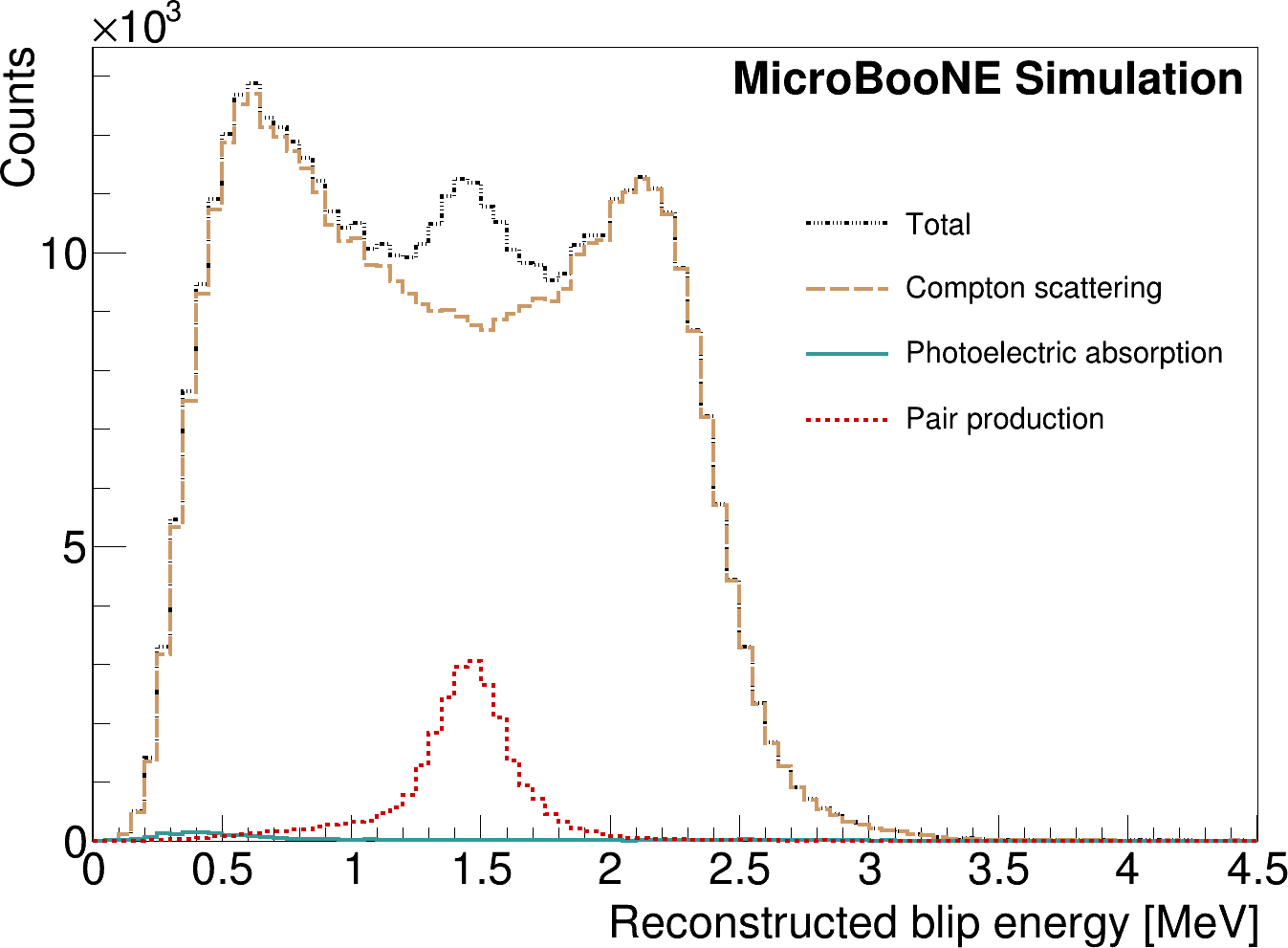}
    \caption{Reconstructed blip energy spectrum in the Monte Carlo simulation sample separated by the blips' originating process. The Monte Carlo simulation primary particles are $2.614$ MeV photons created at the positions of the signal hot spots due to radiological activity in  MicroBooNE's G10 struts, simulating the $\gamma$-rays produced from $^{208}$Tl decays. These are background-subtracted energy distributions of reconstructed blips in the signal region.}
    \label{fig:process}
\end{figure}

 While the energy spectrum of Compton-scattered electrons spreads over a continuum, the pair-production interaction creates an $e^+e^-$ pair with a fixed total energy of $1.592$ MeV, which is $2.614$ MeV minus two times the electron mass. Figure \ref{fig:process} shows the energy distribution of reconstructed blips produced from Compton scattering and pair production of simulated $2.614$ MeV $\gamma$-rays in MicroBooNE. The pair-production peak is well-suited for energy resolution calibration thanks to the monoenergetic nature of its products. This energy signature is made more prominent by the fact that pair production is more probable for the monoenergetic primary $^{208}$Tl decay $\gamma$-ray than for the continuous down-scattered Compton photons, as the pair-production cross section rapidly increases with energy past its $1.02$ MeV threshold \cite{NIST}. 
 
\begin{figure}[h]
        \centering
    \includegraphics[width=0.75\linewidth]{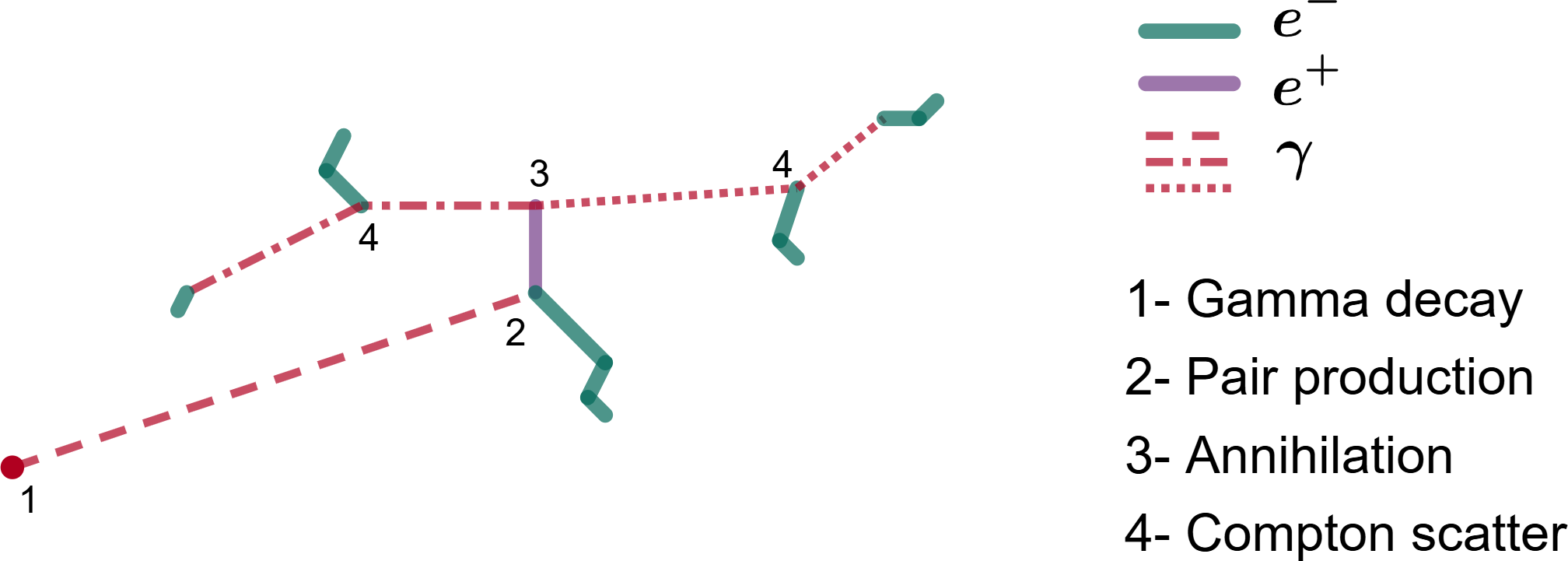}
        \caption{Schematic of an MeV-scale $\gamma$-ray undergoing pair production in liquid argon. The $e^+$ annihilation (`3' in the diagram) yields two back-to-back photons, which Compton scatter or undergo photoelectric absorption (`4'). The alignment of the annihilation products (`4') with the annihilation point (`3') can be used as a topological signature of the pair production (`2').}
        \label{fig:topology}
\end{figure}

 Pair production also differs from Compton scattering in that it produces a very characteristic multi-site event topology. The Compton scattering angle is continuously distributed; moreover, since Compton scattering is the dominant process, one should expect multiple, topologically distinct Compton electrons to appear in the same vicinity in the detector. Meanwhile, the pair-production $e^+$ overwhelmingly annihilates at rest, yielding two back-to-back $0.511$ MeV $\gamma$-rays, which give rise to collinear Compton electrons or photoelectrons on opposite sides of the pair-production point, as illustrated in a cartoon schematic in figure \ref{fig:topology}. Hence, despite MeV-scale $e^+e^-$ pairs being indistinguishable from single electrons in neutrino LArTPCs, the pair production can be identified by the nearly linear topology of the $e^+$ annihilation products.

\section{Radiogenic $^{208}$Tl in the MicroBooNE LArTPC} \label{sec:tl208_uboone}

\begin{figure}[t]
    \centering
    \begin{subfigure}{0.5\textwidth}
        \includegraphics[width=\textwidth]{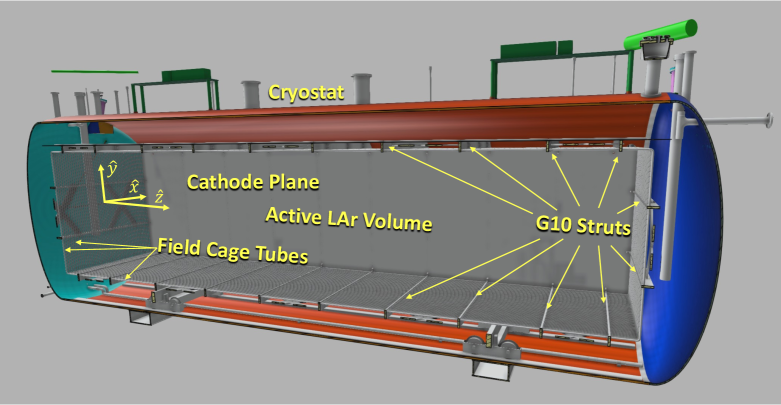}
        \caption{}
        \label{fig:uBooNE}
    \end{subfigure}
     \begin{subfigure}{0.4425\textwidth}\includegraphics[width=\textwidth]{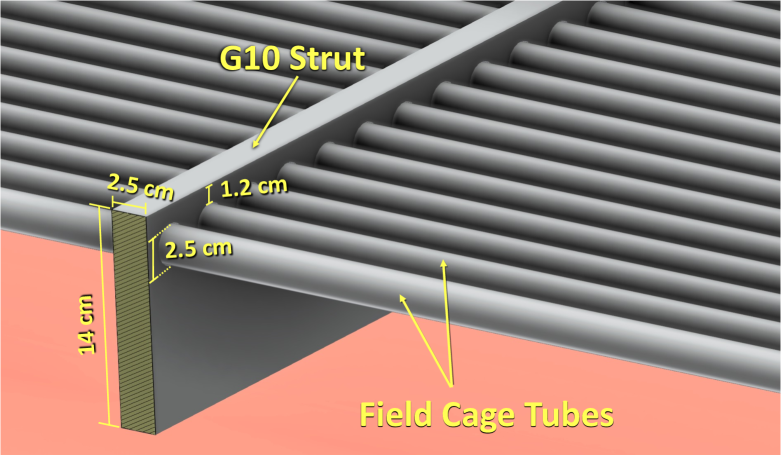}
       \caption{}
       \label{fig:g10_close}
    \end{subfigure}
    \caption{MicroBooNE isometric model: \textbf{a)} Detector view from the anode side. The positions of the cathode plane, field cage tubes, and G10 struts are indicated. \textbf{b)} Close-up of the  interface between the field cage tubes and a G10 strut on the bottom surface of the TPC.}
    \label{fig:g10}
\end{figure}

MicroBooNE (2015--2021) is a LArTPC-based neutrino experiment on Fermilab's Booster Neutrino Beamline (BNB). The $2.56 \times 2.33 \times 10.37$ m$^3$ (w $\times$ h $\times$ l) TPC, depicted in figure \ref{fig:uBooNE}, holds $85$ metric tons of purified LAr in its active volume. Its $yz$-plane faces consist of one cathode plane (at $x=256$ cm) and three anode planes ($3$ mm apart from each other at $x=0$) with a uniform \mbox{$274$ V/m} electric field in between. The active LAr volume is delimited by stainless steel field cage tubes, stabilized with struts composed of G10 fiberglass parallel to the electric field. A thorough description of the MicroBooNE detector can be found in \mbox{ref. \cite{uboone}}. 

The charge collection sensitivity to sub-MeV energy depositions in MicroBooNE is possible due to: argon's low ionization energy ($23.6$ eV per electron); the low residual noise in the wire-readout electronics (equivalent noise charge as low as $300$ $e^-$ \cite{noise_uboone}); the high electron-ion recombination survival probability in the detector's electric field ($60$\% for minimally-ionizing particles); and an electron drift-lifetime ($6.8$--$18.0$ ms \cite{lifetime}) much longer than the time to cross the full drift distance ($2.3$ ms). Each of the two innermost anode planes (`induction' planes) contains $2400$ wires oriented at $\pm60^\circ$ relative to the vertical direction, on which ionization electrons induce bipolar signals as they drift to the `collection' anode plane. The collection plane is made out of $3456$ vertical wires that collect the electrons, registering unipolar pulses \cite{WireCellI}. The wire spacing in all planes is $3$ mm.

It has been demonstrated that the G10 struts contain radioactive impurities, including $^{208}$Tl with a specific activity of $(11.7 \pm 0.2 \text{ (stat) } \pm 3.1 \text{ (syst)})$ Bq/kg \cite{Diego}. These struts, each $15.4$ kg in mass and $2.5$ cm thick, are fabricated from G10, an electrically-insulating epoxied fiberglass laminate \cite{g10}. They are positioned on the bottom and top surfaces of the TPC ($10$ each), as well as the front and back (two each), providing the only mechanical connection between the cathode and anode. Figure \ref{fig:g10_close} shows how the field cage tubes pass through machined holes in the struts, leaving \mbox{$1.2$ cm} of G10 material in the LArTPC active volume. Consequently, MicroBooNE has $^{208}$Tl decay $\gamma$-ray production in the instrumented LAr. This and other G10 radioactive signatures are visible as regions of elevated blip activity (`hot spots') along the edges of the detector.

\section{MicroBooNE datasets and MeV-scale reconstruction} \label{sec:blips}

The analysis described here is performed on a MicroBooNE unbiased dataset of $653{,}367$ event readouts randomly triggered during beam-off periods of MicroBooNE's Run 3 in June--July 2018, corresponding to a cumulative exposure of around $35$ minutes of wall-clock time.  The same data sample was used in previous studies of the presence and attributes of $^{222}$Rn \cite{Radon} and $^{208}$Tl \cite{Diego} in the MicroBooNE TPC.

The pair-production interaction selection cuts (defined in section \ref{sec:selection}) are informed by a Monte Carlo simulation sample of $2.614$ MeV photons generated in the G10 struts and propagated through the detector. MicroBooNE's standard overlay of unbiased beam-off data events adds data-realistic noise to the simulation. An analogous dataset was employed in \mbox{ref. \cite{Diego}}, where the Monte Carlo simulation production with \texttt{LArSoft} \cite{LArSoft} and \texttt{Geant4} \cite{Geant4} tools is further discussed. 

This study is based on reconstructed blips. As mentioned in section \ref{sec:intro}, blips are low-energy, isolated, compact features originated when the wire-based readout system senses mm- to cm-extent ionization clouds created by MeV-scale charged particles. The processing of the datasets follows three steps: raw charge waveform deconvolution, noise filtering, and region-of-interest (ROI) identification with the \texttt{WireCell} toolkit \cite{WireCellI, WireCellII}; `hit' finding and reconstruction with the \texttt{GausHit} algorithm \cite{GausHit}; and `cluster' and blip formation with the custom-built \texttt{LArSoft} toolkit  \texttt{BlipReco} \cite{Radon}. Hits are threshold-crossing pulses in a wire's waveform corresponding to charge collection \cite{WireCellI}. On each plane, hits on adjacent wires at close times are grouped into clusters, which are then matched between planes to form three-dimensional blips. The matching procedure takes into account the clusters' mean time, time span, charge, and wire crossing location to identify clusters on different planes generated by the same drifting electron cloud. Blips may generally be composed of clusters from two or three planes, as long as one of them is on the collection plane, which is the plane used for calorimetry. The blip reconstruction thresholds are lowered by use of looser signal-finding requirements --- ROI forming criteria and minimum hit height --- than those used for most other MicroBooNE analyses. More details on these lowered thresholds are available in \mbox{ref. \cite{Radon}}. 

In this analysis, we consider only three-plane blips to avoid noise and achieve better spatial associations. The $^{208}$Tl $\gamma$-ray pair-production peak lies in the energy region where the blip reconstruction efficiency reaches a maximum of $85$\% ($>1$ MeV) \cite{Radon}. The annihilation-$\gamma$ features, on the other hand, are in a region (< $0.5$ MeV) with $\approx 0$\% reconstruction efficiency for blips, but up to $80$\% cluster reconstruction efficiency on the lower-noise collection plane \cite{Radon}. Therefore, part of the study is carried out with unmatched collection-plane clusters. Furthermore, to avoid potential reconstructed energy loss, we reject blips and clusters next to a dead wire.  The $653{,}367$ events in the data sample comprise $39{,}760{,}791$ total three-plane blips and $309{,}362{,}139$ total collection-plane clusters. The simulation dataset is composed of $71{,}156$ overlaid events, from which we examine the simulated $898{,}058$ three-plane blips and the total (simulated and overlaid) $35{,}021{,}201$ collection-plane clusters.

\section{Pair-production signal selection} \label{sec:selection}

The energy resolution calibration process consists of applying a Gaussian fit to the monoenergetic pair-production peak of $2.614$ MeV $\gamma$-rays from $^{208}$Tl decays. To isolate this feature from other MeV-scale signals, we implement a series of signal selection, noise reduction, and background reduction cuts.

\begin{figure}[h]
    \centering
    \begin{subfigure}{0.55\textwidth}
        \includegraphics[width=\textwidth]{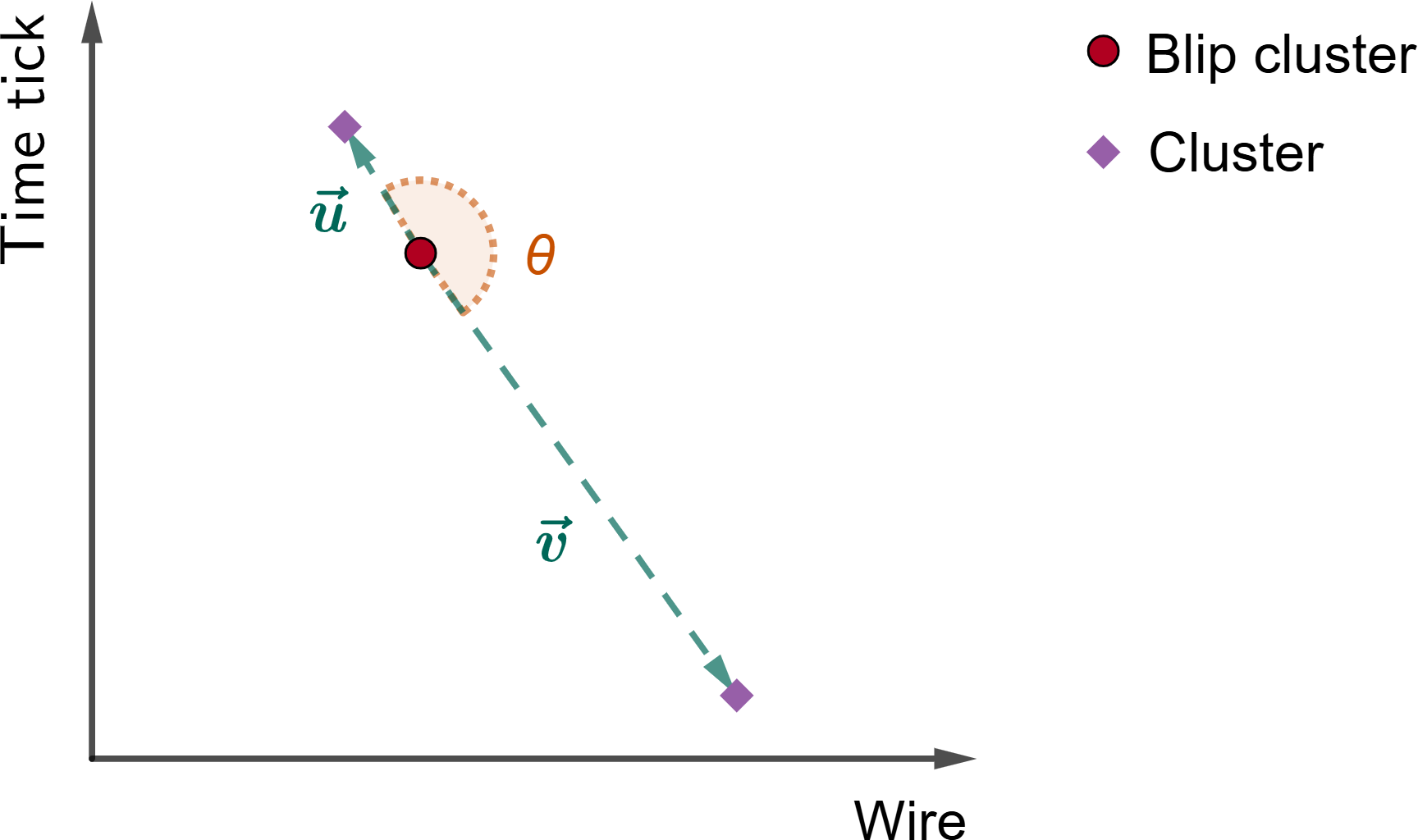}
        \caption{}
    \end{subfigure}
    \hspace{1cm}
    \begin{subfigure}{0.3\textwidth}
        
        \includegraphics[width=\textwidth]{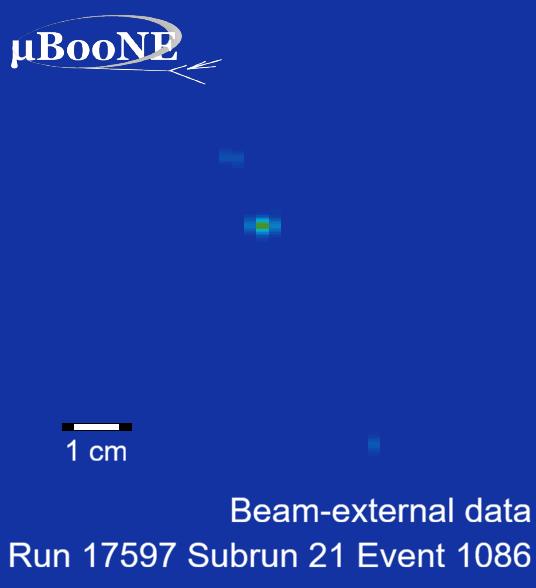}
        \caption{}
    \end{subfigure}

    \caption{Signal topology on the collection plane: the electrons scattered by the annihilation $\gamma$-rays produce clusters that form a $\approx 180^\circ$ angle with respect to the pair-production blip cluster. \textbf{a)} Schematic representation. \textbf{b)} Collection-plane event display fragment containing a selected candidate pair-production blip and its accompanying clusters. }
    \label{fig:topology_cp}
\end{figure}

The pair-production blip candidate selection relies on the $e^+$ annihilation topology depicted in figure \ref{fig:topology}, which is translated into collection-plane clusters in figure \ref{fig:topology_cp}. The annihilation yields two back-to-back photons. Thus, if the energy deposition of the first electron scattered by each annihilation $\gamma$-ray (labeled as `4' in figure \ref{fig:topology}) is reconstructed as a collection-plane cluster, one can draw a nearly-straight line connecting these two clusters to the collection-plane cluster matched into the pair-production blip (to which we refer as the `blip cluster'). Note that the $e^+$ annihilates right after depositing the energy reconstructed in the blip, so there is no significant time or wire separation between the pair-production blip (starting at `2' in figure \ref{fig:topology}) and the annihilation point (`3' in figure \ref{fig:topology}).
   
\begin{figure}[h]
    \centering
    \begin{subfigure}{0.326\textwidth}
        \includegraphics[width=\textwidth]{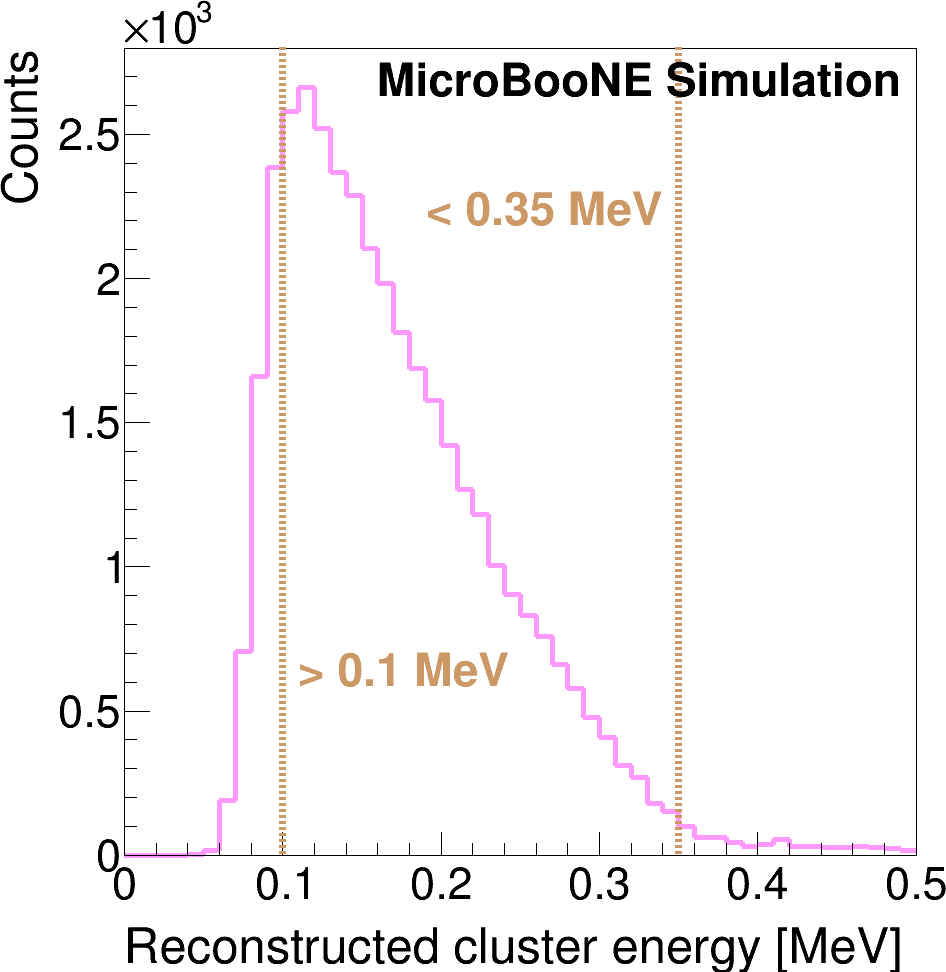}
        \caption{}
    \end{subfigure}
    \hfill
    \begin{subfigure}{0.322\textwidth}
        \includegraphics[width=\textwidth]{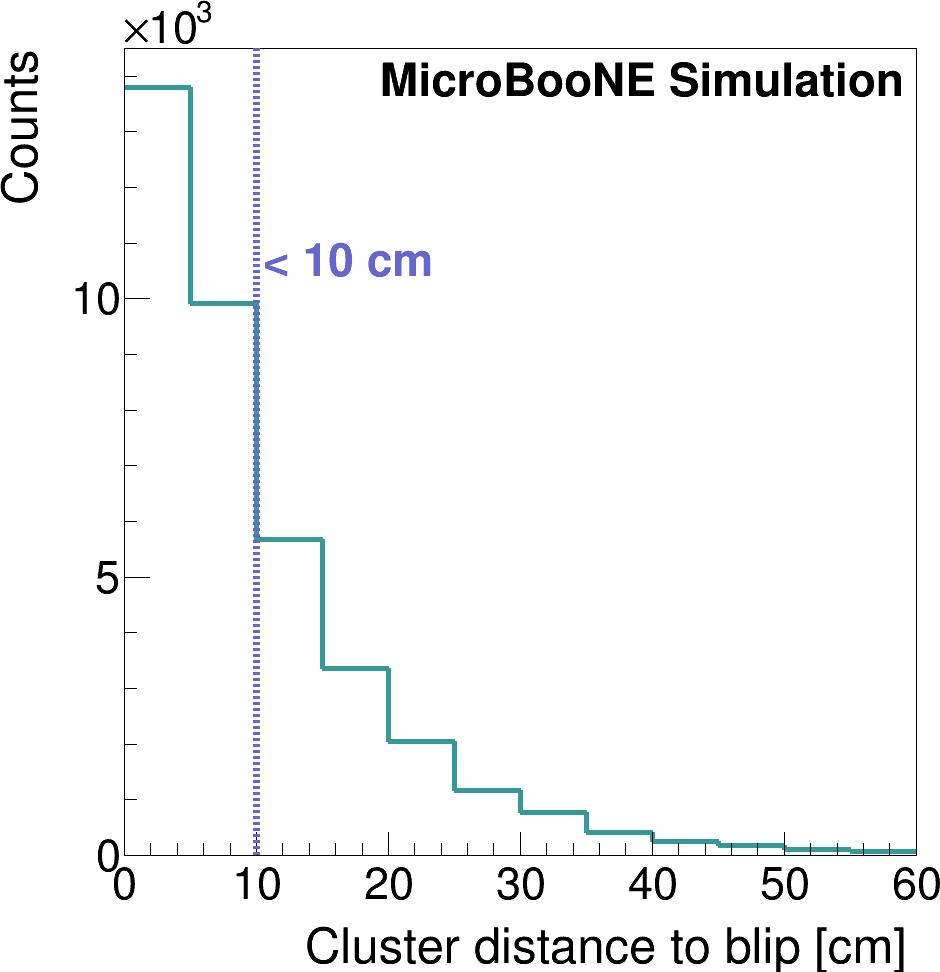}
        \caption{}
    \end{subfigure}
    \hfill
   \begin{subfigure}{0.332\textwidth}
        \includegraphics[width=\textwidth]{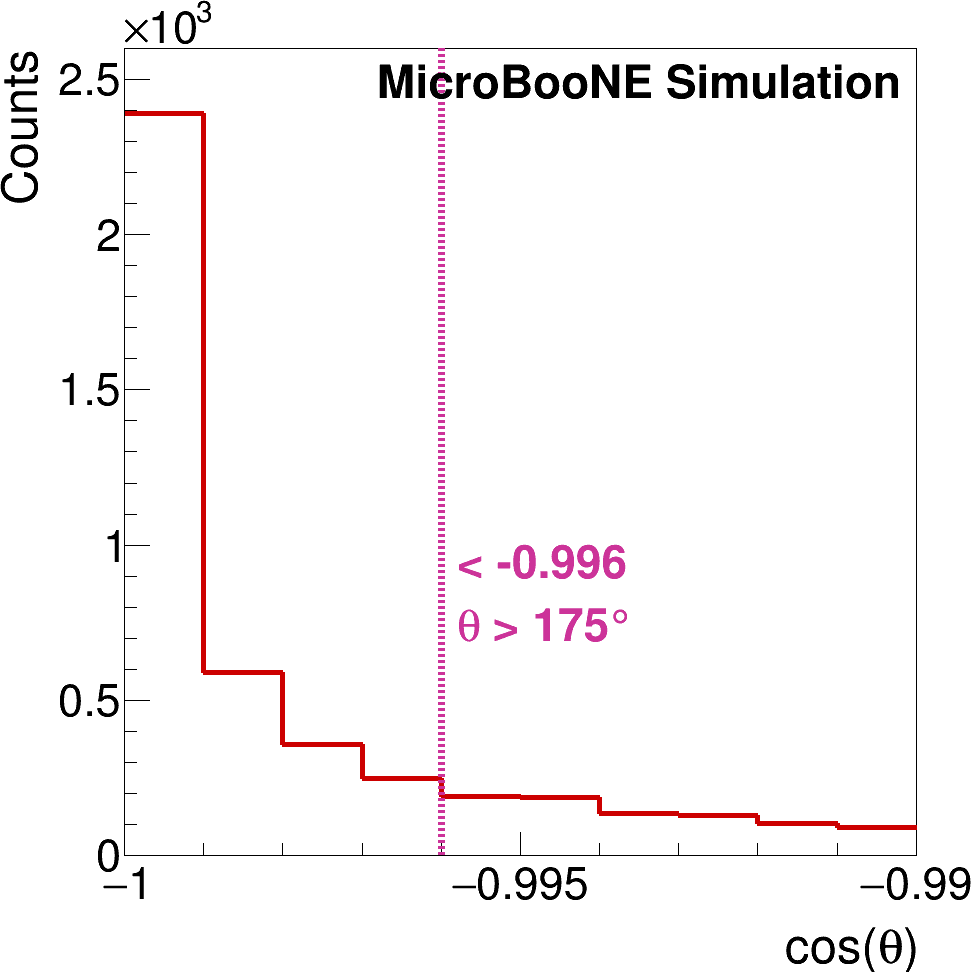}
        \caption{}
    \end{subfigure}
    \caption{True properties of simulated collection-plane clusters composed of charge deposited by particles descending from $\gamma$-rays emitted in pair-production $e^+$ annihilation: reconstructed energy, distance to pair-production blip cluster, and $\cos\theta$, where $\theta$ is the angle between the lines connecting the blip cluster to each collection-plane cluster complying with the energy and distance requirements. These distributions inform the definition of the signal selection cuts. }
    \label{fig:mc_properties}
\end{figure}

We begin the selection with a set of simple signal definition cuts derived from truth-level properties of simulated collection-plane clusters descending from $0.511$ MeV $e^+$-annihilation $\gamma$-rays. Figure \ref{fig:mc_properties} shows the reconstructed attributes of such Monte Carlo simulation collection-plane clusters, as well as the chosen signal selection criteria. We define a pair-production blip candidate as one accompanied by two collection-plane clusters meeting the following requirements: 

\begin{enumerate}[label={(\arabic*)}]
    \item Reconstructed energy $>0.1$ MeV and $<0.35$ MeV. This is consistent with the energy range expected from Compton electrons scattered by $0.511$ MeV photons, with a lower-energy limit imposed by MicroBooNE's collection plane hit-finding threshold.
    \item Less than $10$ cm distant from the blip cluster. This conforms to the attenuation length of $0.511$ MeV photons in LAr, $\approx 9$ cm \cite{NIST}.
    \item The lines connecting each collection-plane cluster to the blip cluster make an angle $\theta$ (see figure \ref{fig:topology_cp}) such that $\cos{\theta}<-0.996$ ($\theta> 175^\circ$). This angle matches the expectation for the products of a two-body decay at rest.
\end{enumerate}
These requirements, relatively tight compared to the full distributions in figure \ref{fig:mc_properties}, are designed to favor purity (rejecting as many non-pair-production blips as possible) over efficiency (selecting all pair-production blips). We hereafter refer to (1), (2), and (3) as the `signal selection cuts'.

 Our analysis is susceptible to noise effects since we do not require that the collection-plane clusters, except for the blip cluster, have matches to other planes. Thus, we apply the following additional cuts aimed at rejecting signal-like topologies that are actually induced by MicroBooNE wire plane noise \cite{noise_uboone}:

\begin{enumerate}[label={(\arabic*)}]
\setcounter{enumi}{3}
\item We reject any pair-production blip candidate with more than two collection-plane clusters satisfying the selection cuts. The existence of several collection-plane clusters aligned with a single blip cluster suggests coincidental alignment in a noisy region of the detector.
\item The time coordinates of the blip cluster and the two aligning collection-plane clusters must differ by more than $3$ time ticks ($1.5$ $\mu$s). This reduces the influence of coherent noise across wires on the final candidate sample.
\item The unmatched collection-plane clusters must not include hits from any wire exhibiting anomalous levels of noise. The noisy-wire numbers are extracted from a list of MicroBooNE noisy channels.

\end{enumerate}

 Moreover, we account for two categories of background: electromagnetic showers and broad-spectrum, cosmogenically-generated $\gamma$-rays. Showers create regions of very high blip density, where the selection requirements may be met by accident. Cosmogenic photons, on the other hand, may undergo true pair-production interactions, generating a broad underlying background in the energy spectrum of selected pair-production blips. The background-reduction cuts, applied in addition to cuts (1) to (6), are as follows:

\begin{enumerate}[label={(\arabic*)}]
    \setcounter{enumi}{6}
    \item Maximum of $275$ blips per event, adding up to a total energy $\leq 350$ MeV. This rejects events with large or high-energy showers.
    \item Candidate blip distance to closest track $>15$ cm (3D distance). This vetoes a large fraction of pair-production interactions of bremsstrahlung photons from cosmic-muon $\delta$ rays.
    \item Total blip energy within $20$ cm (3D distance) of the candidate blip $\leq 1$ MeV. This distance range defines a spherical region around the candidate blip.
    \item Total blip energy within $11$ cm (2D distance on the collection plane) of the candidate blip cluster $\leq 1$ MeV. The distance on the collection plane defines a larger parallelepiped region, hence the smaller required radial acceptance.
    \item No blip with energy $> 3$ MeV within $50$ cm (3D distance) of the candidate blip. This cut has the potential to remove a broad array of electromagnetic backgrounds with minimal impact on the $1.592$ MeV signal, which, by definition, only consists of depositions well below \mbox{$3$ MeV}.
\end{enumerate}

\noindent Cuts (9), (10), and (11) target the extra electromagnetic activity associated with cosmogenically-generated pair-production, absent in pair-production interactions of $^{208}$Tl decay $\gamma$-rays.

\begin{figure}[b]
    \centering
    \begin{subfigure}{0.39\textwidth}
        \includegraphics[width=\textwidth]{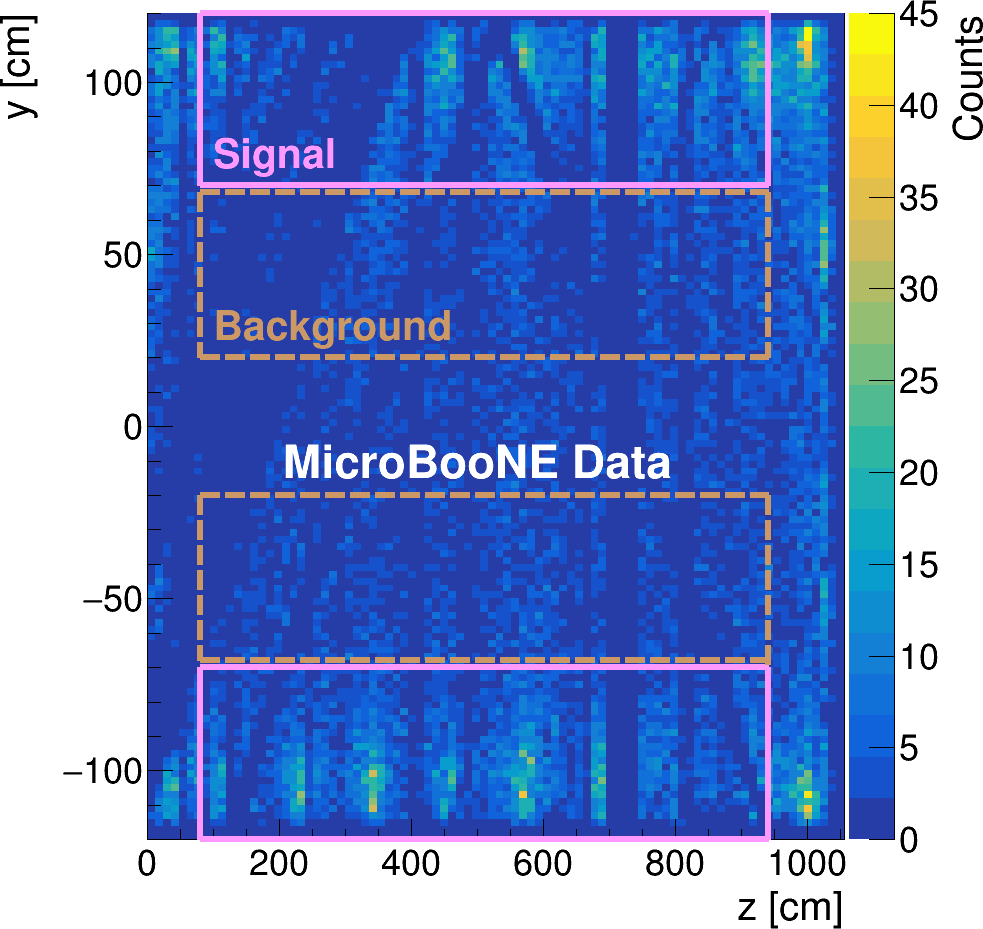}
        \caption{}
        \label{fig:pos}
    \end{subfigure}
    \begin{subfigure}{0.29\textwidth}
        \includegraphics[width=\textwidth]{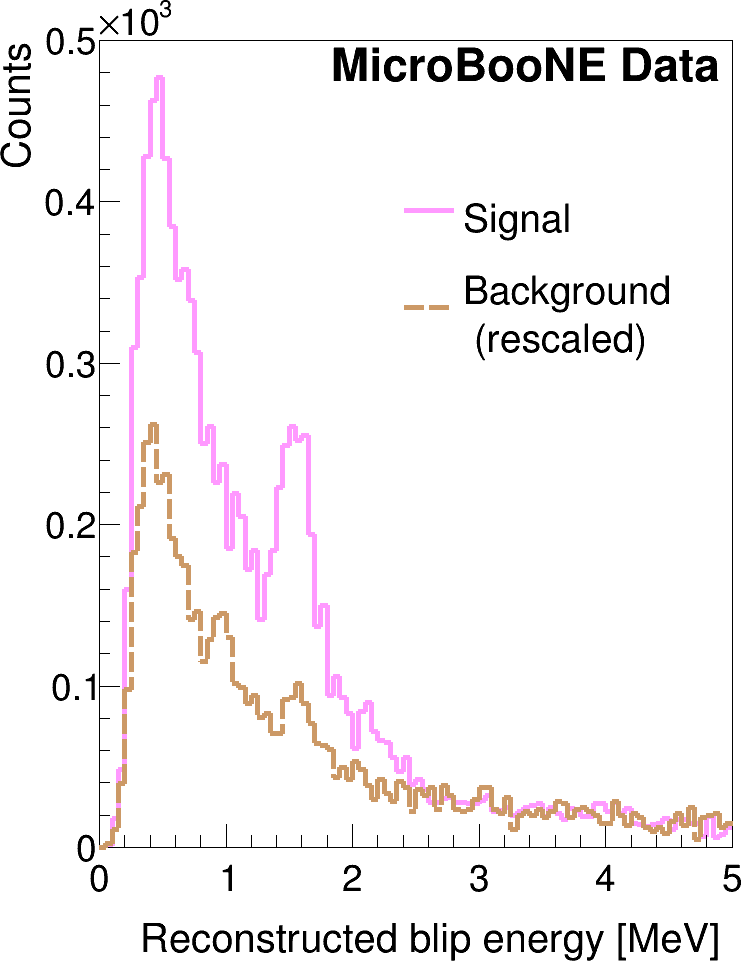}
        \caption{}
        \label{fig:data_bck}
    \end{subfigure}
    \begin{subfigure}{0.29\textwidth}
        \includegraphics[width=\textwidth]{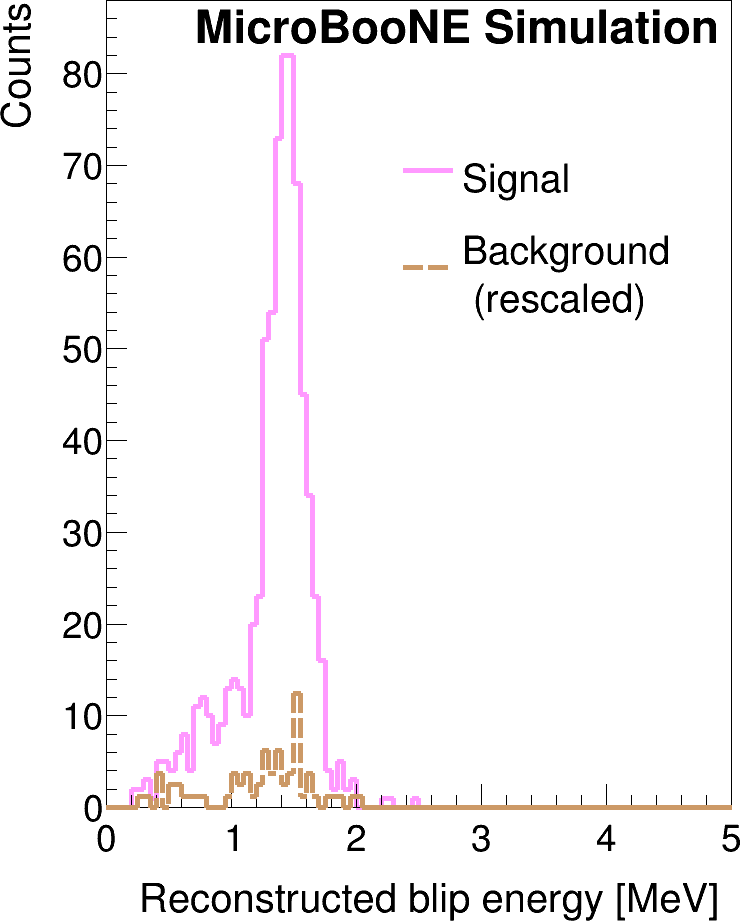}
        \caption{}
        \label{fig:mc_bck}
    \end{subfigure}
    \caption{Signal and background regions. \textbf{a)} $yz$ distribution of $<5$ MeV reconstructed blips in the data sample after all cuts. The signal (background) region used in the analysis is marked in pink (brown). \textbf{b)} (\textbf{c)}) Energy distribution of the selected data (simulation) reconstructed blips in the signal and background regions, following the same color pattern.}
    \label{fig:signal_bck}
\end{figure}

To further reduce the impact of backgrounds on our rare pair-production signal, and increase the fidelity of comparison between data and simulation, we perform a background-subtraction operation based on the signal and background regions indicated in figure~\ref{fig:pos}. The background region, distant from the G10 hot spots enclosed by the signal region, is expected to contain few true pair-production blips from $^{208}$Tl decay $\gamma$-rays, providing a background blip energy distribution that we subtract from the signal. We choose adjacent signal and background regions to account for the $y$ dependence of cosmogenic signals.  Despite this $y$ dependence, we verify consistent resolution results in the top and bottom signal regions, which justifies combining these regions to obtain a higher-statistics signal sample.  The front and back of the detector (low and high $z$, respectively) are excluded as they presented worse resolution, and less well-defined signal versus background differentiation capability, than the top and bottom TPC edges in a simulation comparison. Figure \ref{fig:data_bck} (figure \ref{fig:mc_bck}) displays the energy distributions of data (simulation) blips in the signal and background regions meeting all selection requirements --- (1) to (11). In the data sample, the background is rescaled to match the number of counts in the signal above $2.5$ MeV, since this portion of the spectrum is dominated by cosmogenic signals, which should have roughly the same density in the signal and background regions. This scaling factor ($1.24$) accounts for differences in effective volume between the signal and background regions due to dead wires and variations in blip reconstruction efficiency. The same data-based scaling factor is also applied to the simulation sample background. 

\section{Results and statistical analysis}
\label{sec:analysis}

\begin{figure}[h]
    \centering
    \begin{subfigure}{0.49\textwidth}
        \includegraphics[width=\textwidth]{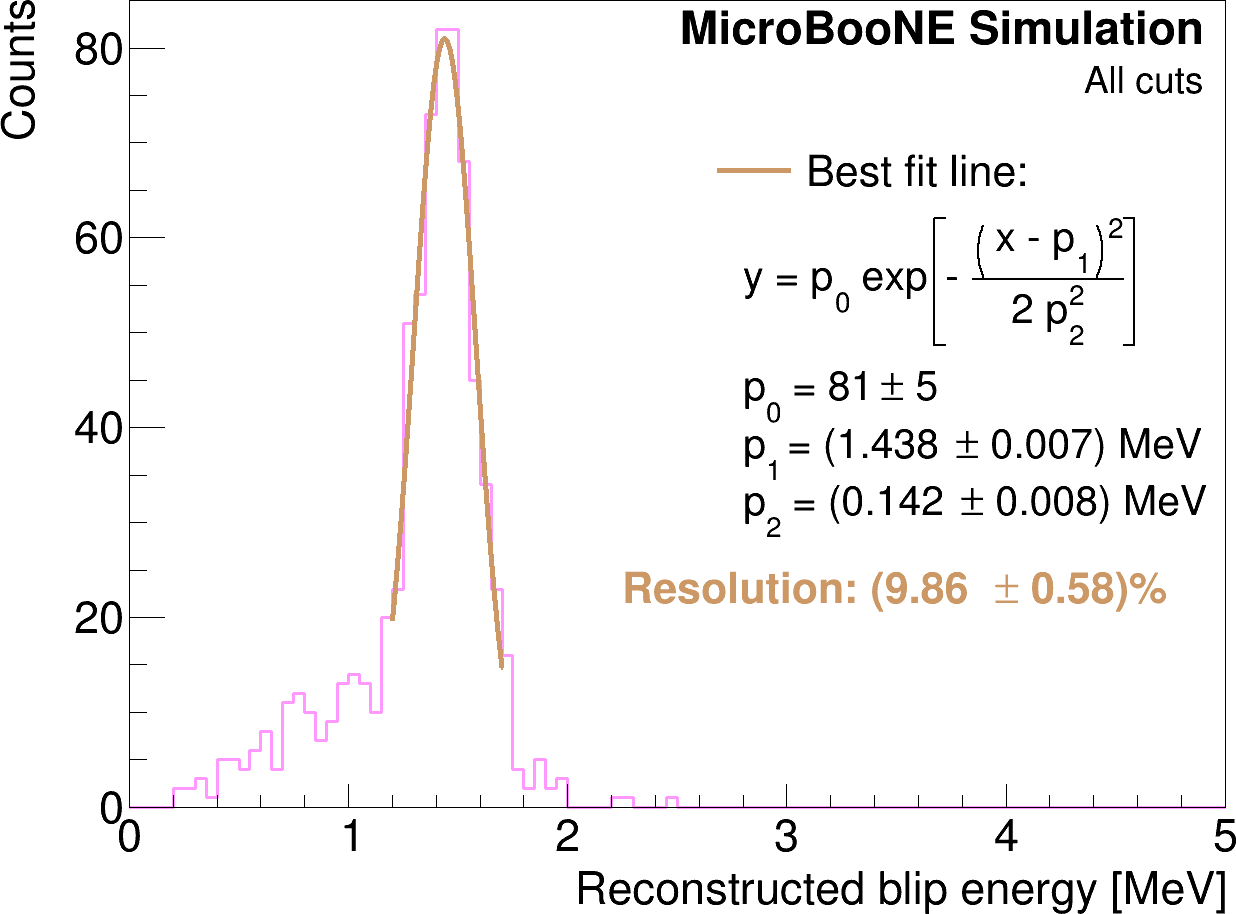}
        \caption{}
        \label{fig:mc_cuts}
   \end{subfigure}
    \begin{subfigure}{0.49\textwidth}
        \includegraphics[width=\textwidth]{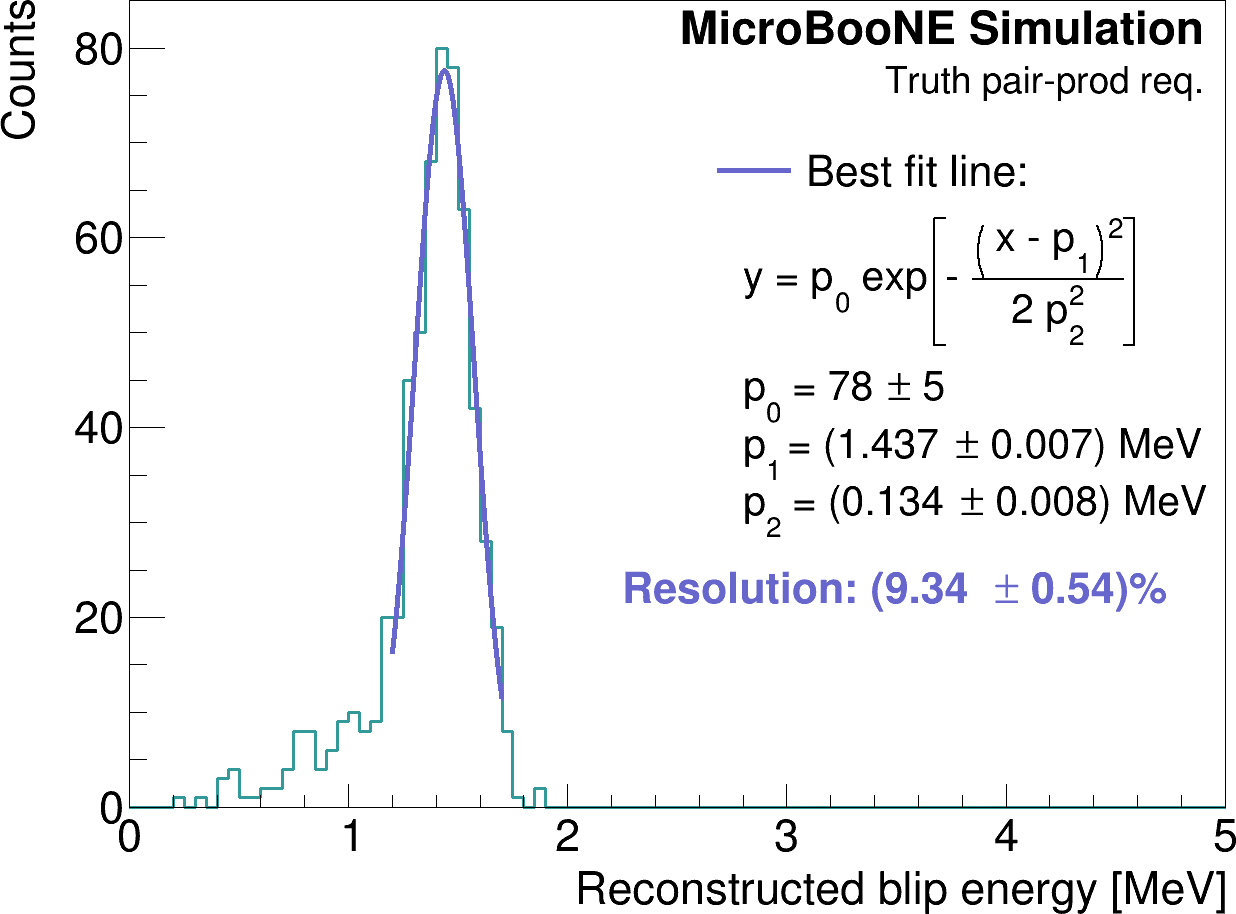}
        \caption{}
        \label{fig:mc_pp}
    \end{subfigure}
    \begin{subfigure}{0.49\textwidth}
        \includegraphics[width=\textwidth]{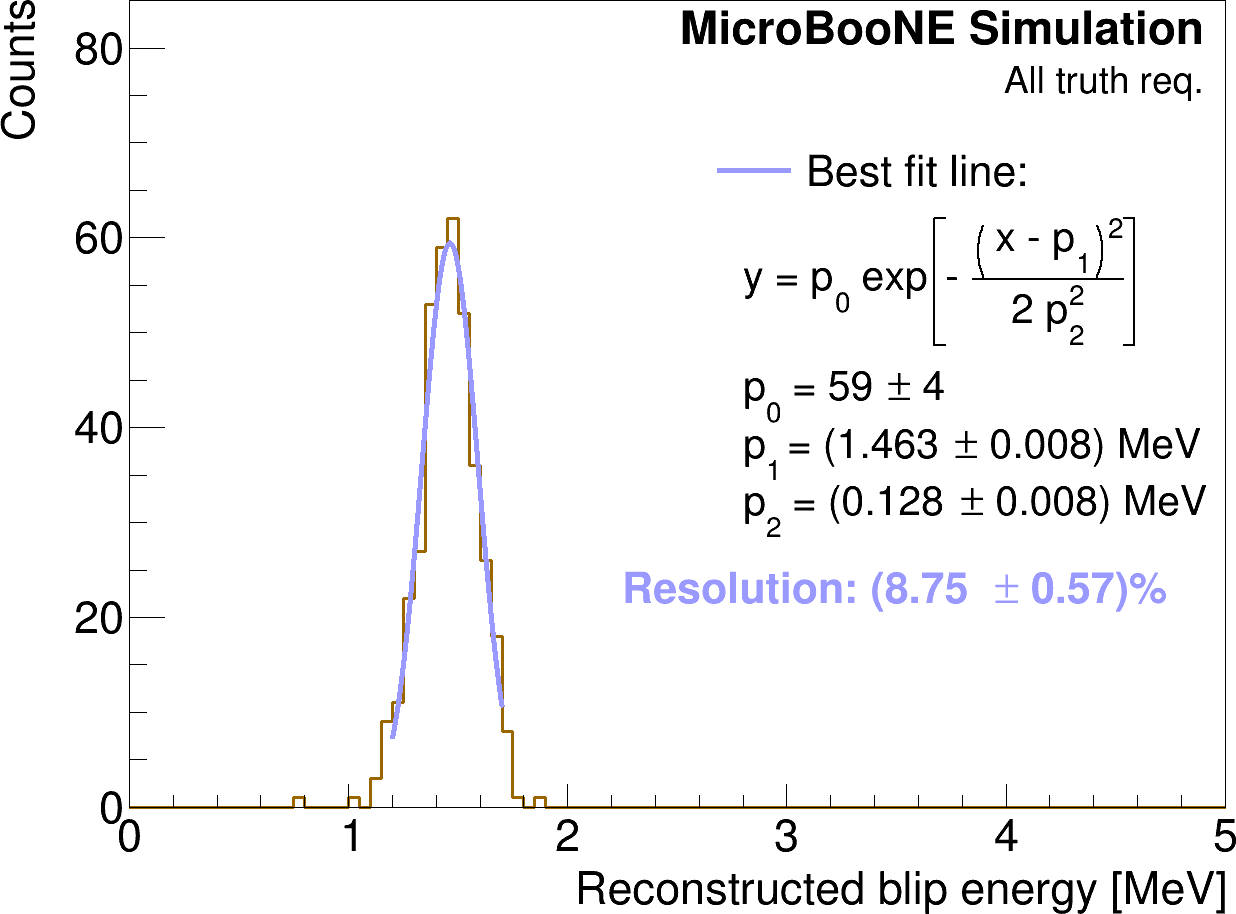}
        \caption{}
        \label{fig:mc_pponly}
    \end{subfigure}
    \begin{subfigure}{0.49\textwidth}
        \includegraphics[width=\textwidth]{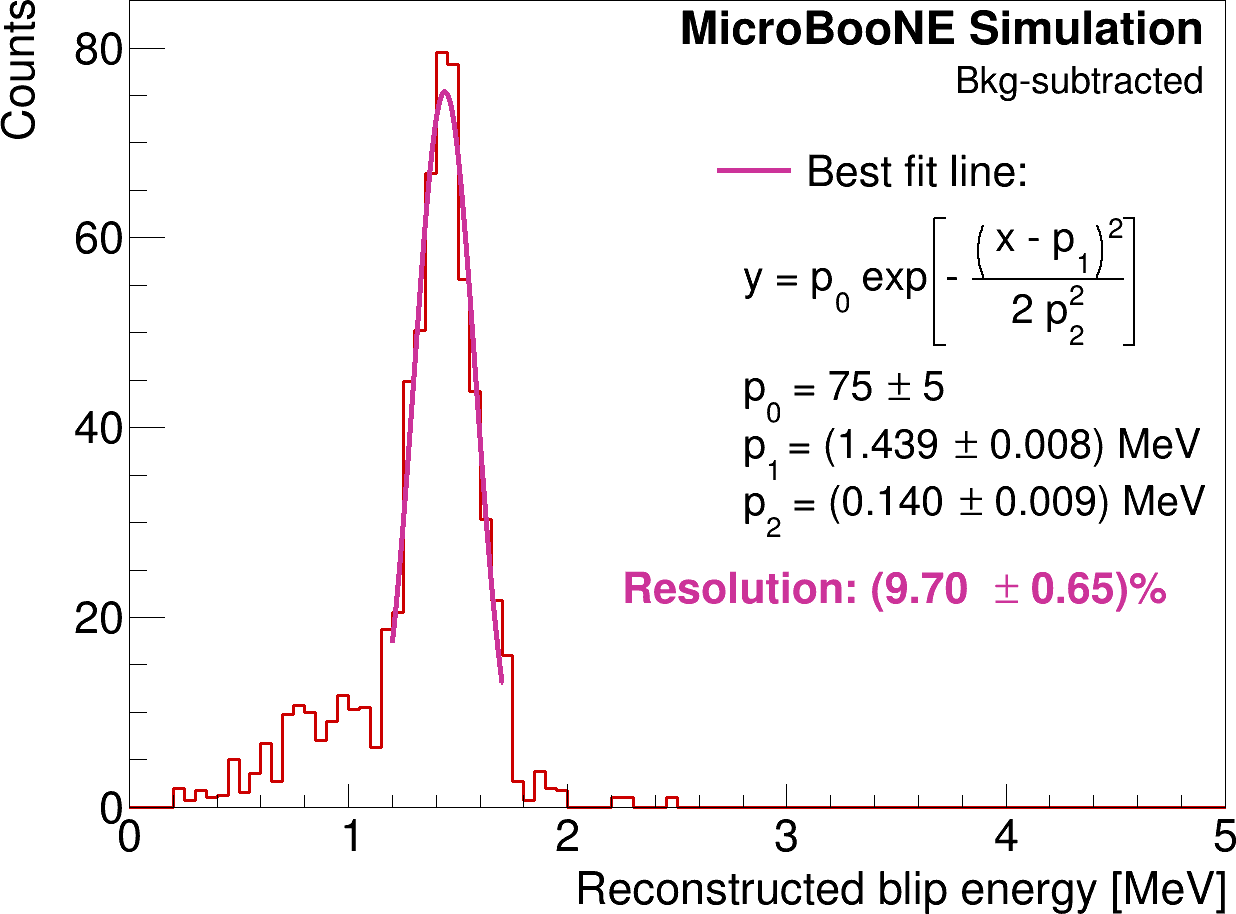}
        \caption{}
        \label{fig:mc_fit}
    \end{subfigure}
    \caption{Energy distribution of Monte Carlo simulation reconstructed blips in the signal region after \textbf{a)} all cuts (signal sample); \textbf{b)} all cuts and truth-information pair-production origin requirement; \textbf{c)} all cuts, pair-production origin requirement, and exclusion of Compton scattering before pair production and bremsstrahlung after pair production; \textbf{d)} all cuts and background subtraction. 
    The fitted Gaussian mean ($p_1$) and standard deviation ($p_2$) of d) define the nominal fractional resolution of the simulation sample.}
    \label{fig:mc_spectra}
\end{figure}

The Monte Carlo simulation blips in the signal region complying with analysis cuts (1)--(11) above (hereafter the simulation `signal sample') form roughly a single peak in energy. Figure \ref{fig:mc_spectra} exhibits the reconstructed blip energy distribution of different subsets of the simulation signal sample. Figures \ref{fig:mc_cuts} and \ref{fig:mc_pp} concern the full sample and the portion of that sample with truth-level pair-production origin, respectively. Figure \ref{fig:mc_pponly} shows a monoenergetic peak obtained by further restricting the truth-information requirement to exclude occurrences that lower the pair-production blip energy: primary $\gamma$-rays Compton scattering before pair-producing and pair-produced particles undergoing radiative losses. Lastly, figure \ref{fig:mc_fit} presents the result of subtracting the energy distribution of the selected blips in the background region from that in the signal region (see figure \ref{fig:mc_bck}), with no truth-level cuts. Its peak ($1.2$--$1.7$ MeV) comprises about $500$ total selected blips.

The blips in the selected simulation signal sample present a $4.9:1$ ratio of pair production to other originating processes, like Compton scattering.  
This high purity is certainly lower in real data due to background sources absent in the simulation, such as cosmogenic signals and blips from other radiogenic $\gamma$-ray lines. 
The selection efficiency is $2.6$\%, which is not surprising given that only $10$\% of the pair-production blips in the simulation dataset are accompanied by one collection-plane cluster created by the first electron scattered by each annihilation $\gamma$-ray. This can be attributed to hit-finding threshold losses as well as to the high probability that one of the annihilation $\gamma$-rays exits the active TPC volume, given that the $^{208}$Tl decay takes place near the edges of the TPC. As previously mentioned, this study prioritizes purity so that the selected blips are as monoenergetic as possible. Overall efficiency is also certainly lower in real data due to the presence of unrelated activity that might cause signal events to be rejected by cuts (4) or (7)--(11).

\begin{figure}[h]
    \centering
    \includegraphics[width=0.75\textwidth]{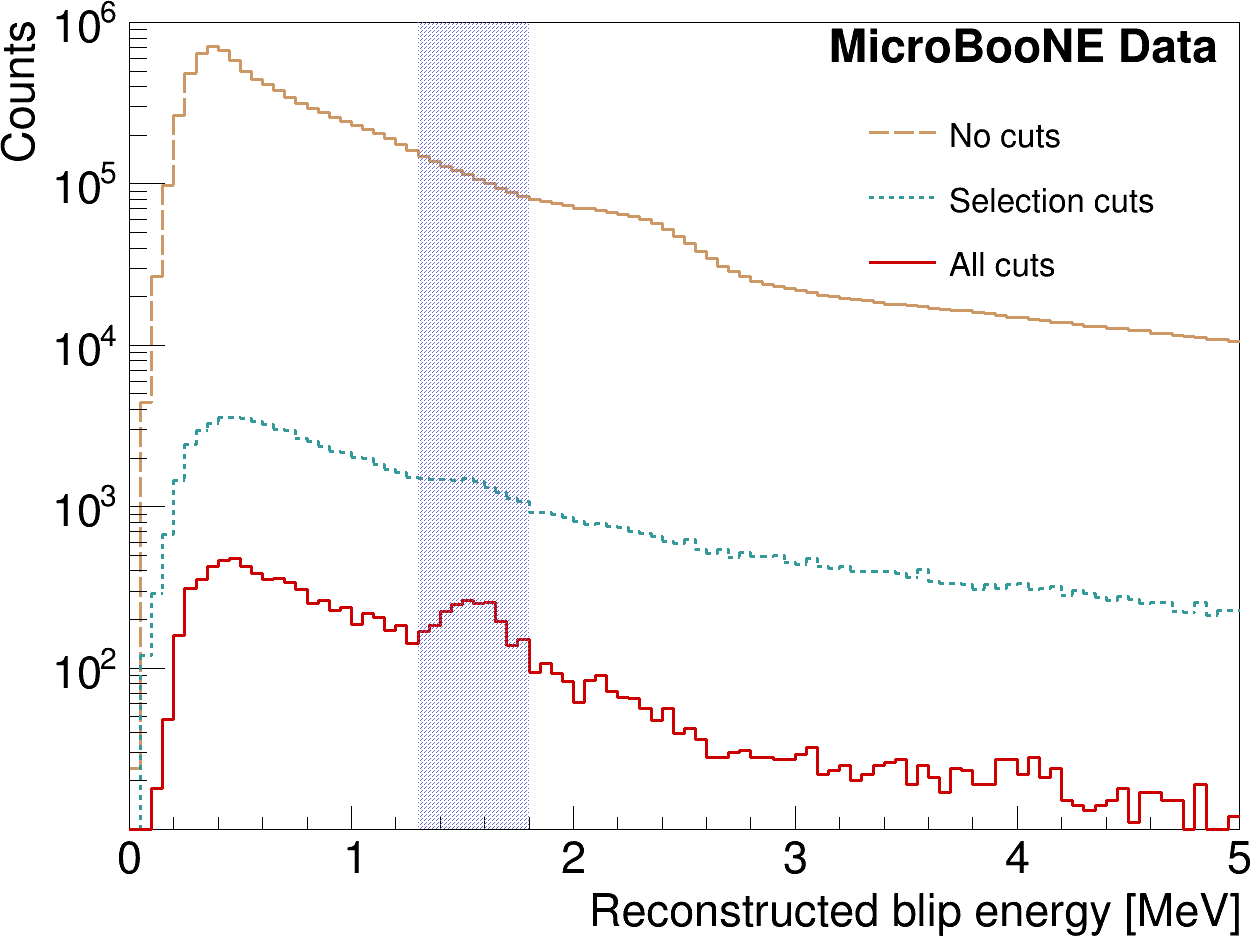}
    \caption{Energy distribution of data reconstructed blips in the signal region: whole spectrum (no cuts), in brown; after selection cuts --- (1), (2), and (3) --- in blue; and after selection, noise-reduction, and background-reduction cuts (signal sample), in red. No background subtraction is applied to these distributions. The peak region ($1.3$--$1.8$ MeV) is highlighted. \vspace{\baselineskip}}
    \label{fig:data_spectra}
\end{figure}

\begin{figure}[h]
    \centering

    \includegraphics[width = 0.75 \textwidth]{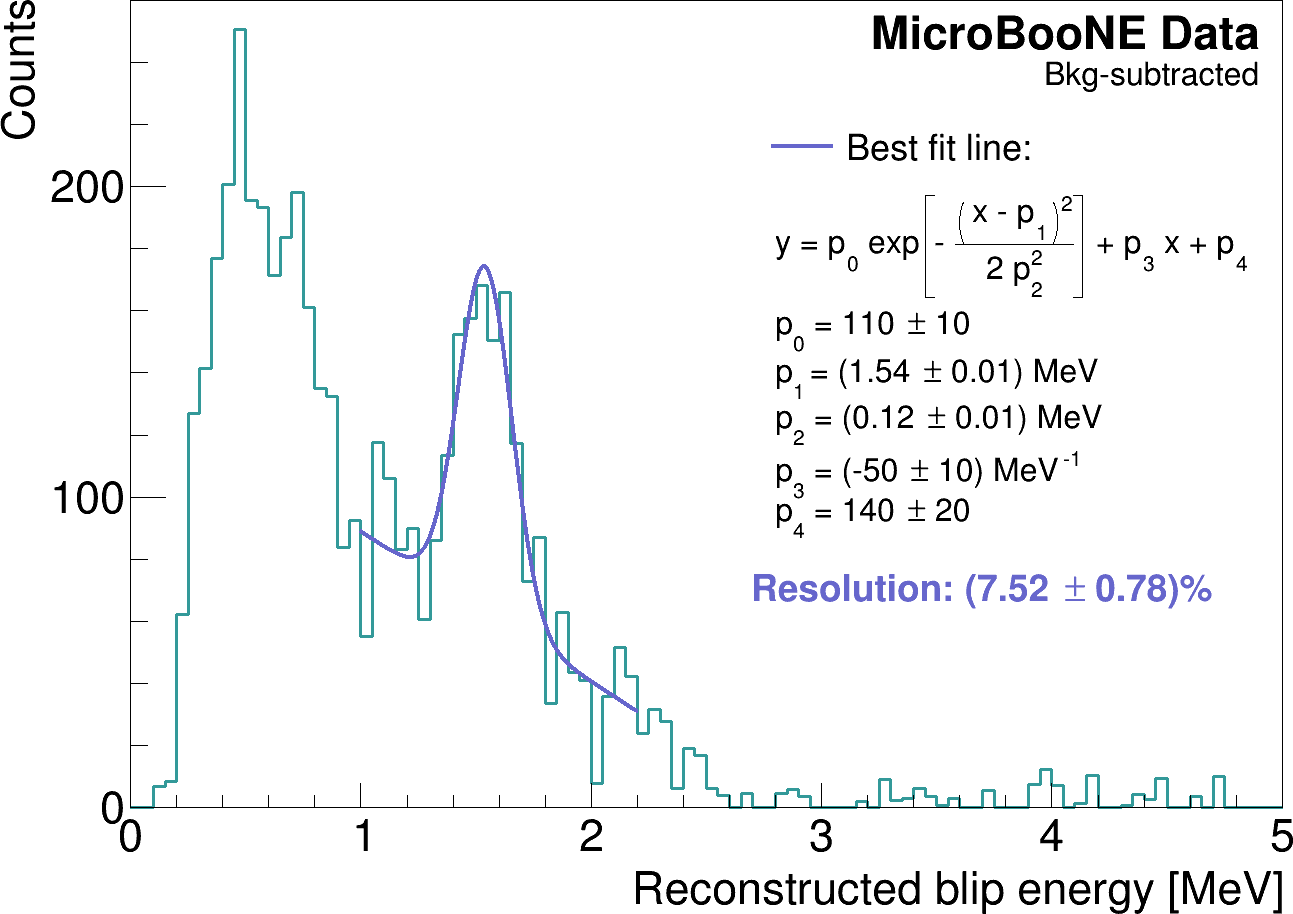}

    \caption{Energy distribution of data reconstructed blips in the signal region after all cuts and background subtraction. The fit function is a Gaussian plus a linear model of the background around the peak. The Gaussian mean ($p_1$) and standard deviation ($p_2$) define the fractional resolution reported in this study. \vspace{-\baselineskip}}
    \label{fig:data_fit}
\end{figure}

 When applied to real data, our signal selection successfully uncovers the $^{208}$Tl decay $\gamma$-ray pair-production signature, as evidenced by the energy distributions of reconstructed blips in the signal region shown in figure \ref{fig:data_spectra}. While the selection cuts alone merely hint at a feature between $1$ and $2$ MeV (blue), the addition of the noise and background reduction cuts reveals a peak between $1.3$ and $1.8$ MeV (red), which is compatible with the predicted $e^+e^-$ pair energy of \mbox{$1.592$ MeV}. After background subtraction (see figure \ref{fig:data_bck}), we obtain the energy spectrum in figure \ref{fig:data_fit}. It contains $1270$ blips in the peak region, some of which are deemed as residual background, since the background feature below $1$ MeV  seems to extend to the peak region.

We apply a fit to the background-subtracted energy distribution in data (figure \ref{fig:data_fit}) to determine the fractional resolution of the peak structure created by the pair-production signature. The fractional resolution is defined as $p_2/p_1$, where $p_1$ is the mean energy and $p_2$ is the standard deviation. The chosen fit function combines a Gaussian with a linear term intended to model the residual background around the peak, and leads to a measured energy resolution of $(7.52 \pm 0.78 \text{ (stat)})\%$. By integrating this fit between $1.3$ and $1.8$ MeV, we estimate the presence of circa $640$ pair-production blip candidates (Gaussian term integral), above a background of circa $630$ blips (linear term integral).  
To determine the magnitude of systematic uncertainty associated with this central-value result, we look for shifts in the fitted resolution that accompany individual changes in the fit range, the assumed background shape, or the signal-to-background ratio. A deviation in the fitted resolution of as much as $0.27$\% is induced by extending the low-energy or high-energy limits of the fit range as low as 0.8 MeV and as high as 2.6 MeV, respectively.  Replacing the linear background shape term with an exponential shape produces a resolution variation of $0.26$\%.  The impact of the magnitude of the background relative to the signal is tested by applying the same fit to the signal region without performing a background subtraction ($0.84$\% change), and by applying it to the full detector region instead of the signal region ($0.60$\%).  We observe substantial deviations in both directions, although they are not symmetric. For a conservative approach, we add in quadrature the largest effect from each of the three systematic categories to obtain a symmetric uncertainty of $\pm$0.92\%.  
This ($7.52 \pm 0.78 \text{ (stat)} \pm 0.92\text{ (syst)}$)\% result is the first reported measurement of MeV-scale energy resolution in a neutrino LArTPC. According to the best-fit parameters of equation (\ref{eq:res}) obtained in \mbox{ref. \cite{Radon}}, energy resolution in the energy region of the pair-production peak is dominated by reconstruction-related systematic effects (i.e., detector energy response non-uniformities, charge deposition on below-threshold wires, etc.). While the present analysis only assesses the resolution at one energy and is therefore not able to estimate the different resolution contributions in equation (\ref{eq:res}), we expect the same reconstruction-related effects to be the major contributor to our resolution result. 

We now evaluate the level of data-simulation agreement of the energy resolution result obtained. The relevant values of $p_1$, $p_2$, and resolution are summarized in table \ref{tab:resolution}. The selected Monte Carlo simulation blip energy distributions in figure \ref{fig:mc_spectra} are fitted with a single Gaussian function given the lack of substantial backgrounds in the simulated 2.614~MeV $\gamma$-ray sample. 
The best-fit resolution results are generally consistent with the $\approx 9$\% found in \mbox{ref. \cite{Radon}} at the same energy, but in the full fiducial volume of the detector. 
Application of background subtraction to the simulation sample does not introduce a significant change in resolution ($9.86$\% to $9.70$\%), presumably because most background sources are not present in the simulation.  
A modest sharpening of resolution can be observed in the full signal sample ($9.86$\%, figure~\ref{fig:mc_cuts}) as one examines only its true pair-production component ($9.34$\%, figure~\ref{fig:mc_pp}) or its truly monoenergetic pair-production component ($8.75$\%, figure~\ref{fig:mc_pponly}).  
This percent-level smearing indicates that the energy-lowering processes discarded in figures ~\ref{fig:mc_pp} and~\ref{fig:mc_pponly} play a sub-dominant role in defining the observed resolution, leaving the monoenergetic nature of the $^{208}$Tl calibration feature largely intact.  
With a full characterization in hand for data and Monte Carlo simulation, we can conclude that this feature's measured energy resolution of ($7.52 \pm 0.78 \text{ (stat)} \pm 0.92\text{ (syst)}$)\% in data and ($9.70 \pm 0.65 \text{ (stat)})$\% in simulation differ by $(2.2\pm1.4)$\%, indicating data-simulation consistency within $1.6\sigma$ significance.  

\begin{table}[h]
    \centering
    \caption{Mean energy ($p_1$), standard deviation ($p_2$), and resolution (main value and statistical uncertainty) obtained from the fits applied to the Monte Carlo simulation and data samples, displayed in figures \ref{fig:mc_spectra} and \ref{fig:data_fit}, respectively. \vspace{0.5\baselineskip}}
    \begin{tabular}{c|c|c|c} \textbf{Sample} & \textbf{Mean} (MeV) & \textbf{Standard}  & \textbf{Resolution} (\%)
    \\ & & \textbf{Deviation} (MeV)   \\
     \hline
     \hline
        \multicolumn{4}{c}{Monte Carlo} \\
     \hline
       All cuts (1)--(11) & $1.438 \pm 0.007$ & $0.142 \pm 0.008$ & $\mathbf{9.86\pm0.58}$ \\
       Truth pair-production requirement & $1.437 \pm 0.007$ & $0.134 \pm 0.008$ & $\mathbf{9.34\pm0.54}$\\
       All truth requirements  & $1.463 \pm 0.008$ & $0.128 \pm 0.008$ & $\mathbf{8.75\pm0.57}$ \\
       Background subtracted & $1.439 \pm 0.008$ & $0.140 \pm 0.009$ & 
       $\mathbf{9.70\pm0.65}$ \\
       \hline
       \hline
       \multicolumn{4}{c}{Data}\\
       \hline
       Background subtracted & $1.54 \pm 0.01$ & $0.12 \pm 0.01$ & 
       $\mathbf{7.52\pm0.78}$ \\
       \hline
    \end{tabular}
    \label{tab:resolution}
\end{table}

We note that the fit and error propagation methods used to provide the energy resolution results described above also deliver a mean energy peak value for the Gaussian pair-production feature and an associated uncertainty. 
Best-fit peak values of $(1.54 \pm 0.01 \text{ (stat)} \pm 0.02\text{ (syst)}$)~MeV and $(1.439 \pm 0.008 \text{ (stat)})$~MeV are determined for the data and simulation samples, respectively.  
Percent-level agreement between fitted reconstructed energy values and the true expected energy value of $1.592$ MeV should not be expected given the detector response issues in play in relating true energy deposited to a reconstructed energy, such as the assumption of a linear ionization charge to energy conversion.  
The data-simulation offset in peak location, $(7.2 \pm 1.0 \text{ (stat)} \pm 1.5\text{ (syst)}$)\%, is more surprising, given a previously reported value of $(3.1 \pm 0.2 \text{ (stat)} \pm 1.2\text{ (syst)})$\% from the same $^{208}$Tl decay $\gamma$-ray's Compton edge at $\approx$ 2.4~MeV.  
By adjusting and redoing the Compton edge energy scale fit procedure of \mbox{ref. \cite{Diego}}, we confirm that this $\approx$ 2$\sigma$ difference is not due to distinct signal and background region definitions between the two studies. Still, it should be noted that both studies are restricted to the edges of the detector, where the response is affected the most by factors such as space charge effects and electric field variations.
If sub-percent-precision energy scale calibrations are desired for MeV-scale physics measurements in neutrino LArTPCs, higher-statistics assessments of the $^{208}$Tl Compton edge and pair-production peak features should be performed in future studies to more precisely probe the nature of the observed data-simulation offsets. 
In addition to higher statistics, future analyses could improve upon these results by systematically probing the position dependence of MeV energy scales, by incorporating more complete models of the entire energy spectrum of radiogenic activity, and by continuing to expand the catalog of identified radiogenic and cosmic spectral features.  


\section{Summary} \label{sec:summary}
We report a new MeV-scale selection in MicroBooNE that identifies a clear but rare nearly monoenergetic spectral feature from pair-production interactions of $2.614$ MeV $\gamma$-rays generated by ambient $^{208}$Tl decays. This analysis uniquely leverages the impressively low energy thresholds ($100$ keV-scale) achieved for MicroBooNE's collection-plane signals.

Applying this selection to a beam-off MicroBooNE dataset of $653{,}367$ events containing $39{,}760{,}791$ total blips, we identify $640$ signal blip candidates above a similarly sized background contribution.  The selected events reveal a peak in the reconstructed blip energy distribution centered at $(1.54 \pm 0.01 \text{ (stat)} \pm 0.02\text{ (syst)}$)~MeV that has a Gaussian resolution of ($7.52 \pm 0.78 \text{ (stat)} \pm 0.92\text{ (syst)}$)\%. A Monte Carlo simulation sample of $2.614$ MeV $\gamma$-rays in MicroBooNE is also examined, yielding a reconstructed energy resolution of ($9.70 \pm 0.65$ \text{ (stat)})\% for the pair-production peak, centered at ($1.439\pm0.008$) MeV. The resolution values obtained from data and simulation are consistent. 

This is the first direct characterization of MeV-scale energy resolution in a neutrino LArTPC and is a promising result in the context of achieving sensitivity to supernova and solar neutrinos in LArTPC-based experiments.  
Additional investigations with higher-statistics samples should be carried out to more precisely investigate potential differences in energy scale and resolution between data and Monte Carlo simulation.  
Future studies similar to the one described here may provide a useful template procedure for performing neutrino LArTPC calibrations in the absence of large through-going muon fluxes. The success of such calibrations using blips will depend on the presence of radioactive isotopes in the detector structure, as well as on low-energy sensitivity and reconstruction thresholds comparable to MicroBooNE’s.

\section*{Acknowledgments}

This document was prepared by the MicroBooNE collaboration using the resources of the Fermi National Accelerator Laboratory (Fermilab), a U.S. Department of Energy, Office of Science, Office of High Energy Physics HEP User Facility. Fermilab is managed by Fermi Forward Discovery Group, LLC, acting under Contract No. 89243024CSC000002. MicroBooNE is supported by the
following: 
the U.S. Department of Energy, Office of Science, Offices of High Energy Physics and Nuclear Physics; 
the U.S. National Science Foundation; 
the Swiss National Science Foundation; 
the Science and Technology Facilities Council (STFC), part of United Kingdom Research and Innovation (UKRI);
the Royal Society (United Kingdom);
the UKRI Future Leaders Fellowship;
the NSF AI Institute for Artificial Intelligence and Fundamental Interactions;
and the European Union’s Horizon 2020 research and innovation programme under the Marie Sk\l{}odowska-Curie grant agreement No. 101003460 (PROBES). Additional support for the laser calibration system and cosmic ray tagger was provided by the Albert Einstein Center for Fundamental Physics, Bern, Switzerland. We also acknowledge the contributions of technical and scientific staff to the design, construction, and operation of the MicroBooNE detector as well as the contributions of past collaborators to the development of MicroBooNE analyses, without whom this work would not have been possible. For the purpose of open access, the authors have applied a Creative Commons Attribution (CC BY) public copyright license to any Author Accepted Manuscript version arising from this submission.

\bibliographystyle{JHEP}
\bibliography{bibliography}

\end{document}